\documentclass[twocolumn]{aa} 
\usepackage{graphicx,longtable}
\usepackage[figuresright]{rotating}

\usepackage[varg]{txfonts}
\usepackage{epsfig}
\def\c{(c)}

\def\mincir{\raise -2.truept\hbox{\rlap{\hbox{$\sim$}}\raise5.truept
\hbox{$<$}\ }}
\def\magcir{\raise -2.truept\hbox{\rlap{\hbox{$\sim$}}\raise5.truept
\hbox{$>$}\ }}
\begin{document}

\title{An active state of the BL Lac Object Markarian 421 detected by {\it INTEGRAL} in April 2013\thanks{Based on observations obtained with \emph{INTEGRAL}, an 
ESA mission with instruments and science data centre funded by ESA member 
states (especially the PI countries: Denmark, France, Germany, Italy, 
Switzerland, Spain, Czech Republic and Poland), and with the participation 
of Russia and the USA.}}

\author{E. Pian\inst{1,2,3}
\and
M. T\"urler\inst{4}
\and
M. Fiocchi\inst{5}
\and
R. Boissay\inst{4}
\and
A. Bazzano\inst{5}
\and
L. Foschini\inst{6}
\and
F. Tavecchio\inst{6}
\and
V. Bianchin\inst{1}
\and
G. Castignani\inst{7}
\and
C. Ferrigno\inst{4}
\and
C.M. Raiteri\inst{8}
\and
M. Villata\inst{8}
\and
V. Beckmann\inst{9}
\and
F. D'Ammando\inst{10,11,12}
\and
R. Hudec\inst{13,14}
\and
G. Malaguti\inst{1}
\and
L. Maraschi\inst{15}
\and
T. Pursimo\inst{16}
\and
P. Romano\inst{17}
\and
S. Soldi\inst{18}
\and
A. Stamerra\inst{3}
\and
A. Treves\inst{19}
\and
P. Ubertini\inst{5}
\and
S. Vercellone\inst{17}
\and
R. Walter\inst{4}
}

\institute{
INAF, Istituto di Astrofisica Spaziale e Fisica Cosmica, via P. Gobetti 101,  40129 Bologna, Italy
              \email{elena.pian@sns.it}
\and
Scuola Normale Superiore, Piazza dei Cavalieri 7,  56122 Pisa, Italy
\and
INFN, Sezione di Pisa, Largo Pontecorvo 3,  56127 Pisa, Italy
\and
ISDC, Geneva Observatory,  University of Geneva, Chemin d'Ecogia 16, 1290, Versoix, Switzerland
\and
Istituto di Astrofisica e Planetologia Spaziali, Via Fosso del Cavaliere 100,  00133 Roma, Italy
\and
INAF-Osservatorio Astronomico di Brera, Via Bianchi 46, 23207 Merate (LC), Italy
\and
SISSA-ISAS, Via Bonomea 265, 34136, Trieste, Italy
\and
INAF-Osservatorio Astronomico di Torino, Strada Osservatorio 20, 10025 Pino Torinese (TO), Italy
\and
Fran\c{c}ois Arago Centre,  APC, Universit\'e Paris Diderot, CNRS/IN2P3, 10 rue Alice Domon et L\'eonie Duquet,
75205 Paris Cedex 13, France
\and
Dipartimento di Fisica, Universit\`a degli Studi di Perugia, Via A. Pascoli, 06123 Perugia, Italy
\and
INFN, Sezione di Perugia, Via A. Pascoli, 06123 Perugia, Italy
\and
INAF, Istituto di Radioastronomia, Via P. Gobetti 101, 40129 Bologna, Italy
\and
Astronomical Institute, Academy of Sciences, Fricova 298, 25165 Ondrejov, Czech Republic
\and
Czech Technical University in Prague, Faculty of Electrical Engineering, Czech Republic
\and
INAF-Osservatorio Astronomico di Brera,  Via Brera 28, 20100 Milano, Italy
\and
Nordic Optical Telescope, Apartado 474, 38700 Santa Cruz de La Palma, Spain
\and
INAF, Istituto di Astrofisica Spaziale e Fisica Cosmica, via U. La Malfa 153, 90146 Palermo, Italy
\and
APC, Universit\'e Paris Diderot, CNRS/IN2P3, 10 rue Alice Domon et L\'eonie Duquet, 75025 Paris Cedex 13, France
\and
Universit\`a  degli Studi dell'Insubria, Via Valleggio 11, 22100 Como, Italy
}

   \date{}

  \abstract
   {}
   {Multiwavelength variability of blazars offers indirect, but very effective, insight into their powerful engines and on the mechanisms through which energy is propagated from the centre down the jet.   The BL Lac object Mkn~421 is a TeV emitter, a bright blazar at all wavelengths, and therefore an excellent  target for variability studies.}
   {We activated \emph{INTEGRAL} observations of Mkn~421  in an active state on 16-21 April 2013, and complemented them with {\it Fermi}-LAT data.}
   {We obtained well sampled optical, soft, and hard X-ray light curves that show the presence of two flares and  time-resolved spectra in the 3.5-60 keV (JEM-X and IBIS/ISGRI) and 0.1-100 GeV ({\it Fermi}-LAT) ranges.      The average flux in the 20-100 keV range is $9.1 \times 10^{-11}$ erg s$^{-1}$ cm$^{-2}$ ($\sim$4.5 mCrab) and the nuclear average apparent magnitude, corrected  for Galactic  extinction, is $V \simeq 12.2$.  In the time-resolved X-ray spectra, which are described by broken power laws and, marginally better, by log-parabolic laws, we see a hardening  that  correlates with flux increase, as expected in refreshed energy injections in a population of electrons that later cool  via synchrotron radiation.   The hardness ratios between the JEM-X fluxes in two different bands and between the JEM-X and IBIS/ISGRI fluxes confirm this trend.  
During the observation, the variability level increases monotonically from the optical to the hard X-rays, while the large LAT errors do not allow a significant assessment of the MeV-GeV variability.   The cross-correlation analysis during the onset of the most prominent flare suggests a monotonically increasing delay of the lower frequency emission with respect to that at higher frequency, with a maximum time-lag of about 70 minutes, that is however not well constrained.   The  spectral energy distributions from the optical to the TeV domain were compared to homogeneous models of blazar emission based on synchrotron radiation and synchrotron self-Compton scattering.  They represent a satisfactory description, except in the state corresponding to the LAT softest spectrum and highest  flux.}
{Multiwavelength variability  of Mkn~421 can be very complex, with patterns changing from epoch to epoch down to intra-day timescales, depending on the emission state. This makes accurate  monitoring of this source  during bright hard X-ray states necessary and calls for the elaboration of multicomponent, multizone,  time-dependent models.}
   

\authorrunning{E.Pian et al.}
\titlerunning{{\it INTEGRAL} observations of Mkn~421}

   \maketitle
%

\section{Introduction}

Blazars are active multiwavelength  extragalactic sources, with compact  inner engines powered by supermassive black holes and with relativistic jets pointing at small angles with respect to the observer.  They are the most luminous persistent sources in the Universe, reaching bolometric luminosities as large as $10^{49}$ erg/s, often dominated 
by the gamma-ray output (Ghisellini et al. 2011), and are highly variable, with doubling timescales ranging from seconds to years.  Recent dense and accurate monitorings have shown that, while in general the variation amplitudes are higher at  shorter wavelengths (e.g. Pian et al. 2011), the  multiwavelength behaviour is  often more complex and calls into question the role of multiple radiation components (Aleksi{\'c} et al. 2012; Bonning et al. 2012; D'Ammando et al. 2013).  The blazar spectrum is characterised by a thin synchrotron component that peaks, in a $\nu f_\nu$ representation, in 
the wavelength range from far-infrared to X-rays, and by an inverse Compton component that extends from X-rays to the TeV range, peaking at MeV-GeV  energies (Falomo, Pian, \& Treves 2014).  
These characteristic frequencies  vary in an anti-correlated way with respect to bolometric luminosity  (Fossati et al. 1998;  see however Padovani, Giommi \& Rau 2012), leading to a broad distinction of blazars into low-frequency-peaked and high-frequency-peaked objects (Padovani \& Giommi 1995).
Assuming that the emitting region responsible for the emission at X-ray and gamma-ray energies is homogeneous, the ratio between the peak  frequencies of the two components represents the characteristic cooling energy of the relativistic particles in the jet plasma (barring Klein-Nishina suppression).

Mkn~421 ($z = 0.031$) is  a known bright BL Lac object (Ulrich 1973; Colla et al. 1975) of the high-frequency-peaked type, a strong and variable X-ray source up to the hard X-rays, and belongs to the complete sample
of X-ray blazars selected above 15 keV by the {\it Swift}/Burst Alert Telescope  survey (Ajello et al. 2009).    The absence of strong optical  emission lines indicates 
that photon sources external to the jet, e.g. an accretion disk or a dust torus, are modest and do not contribute significantly to the electron 
cooling.  As a consequence, the synchrotron component  peaks at relatively high frequencies also during quiescence (soft-X-rays) and the synchrotron photons are the 
main targets for Compton up-scattering by the relativistic particles.  This synchrotron self-Compton component peaks  at very high energies, 
making this blazar a strong TeV emitter, although the TeV spectrum is often heavily suppressed by the Klein-Nishina effect (Maraschi et al. 1999; Fossati et al. 2008; Mankuzhiyil et al. 2011).
Mkn~421 was the target of many multiwavelength campaigns involving high energy satellites, Cherenkov telescopes and ground-based optical and radio facilities (Takahashi et al. 1996; Maraschi et al. 1999; Malizia et al. 2000;  Fossati et al. 2000a,b; Brinkmann et al. 2001;  Albert et al. 2007a; Lichti et al. 2008; Fossati et al. 2008; Donnarumma et al. 2009; Acciari et al. 2009;  Ushio et al. 2009; Tramacere et al. 2009; Horan et al. 2009; Aleksi{\'c} et al. 2010; Isobe et al. 2010; Abdo et al. 2011; Acciari et al. 2011; Barres de Almeida 2011; Aleksi{\'c} et al. 2012; Shukla et al. 2012).  
The variations at X-ray and TeV frequencies are often very well correlated with   no measurable delay (Maraschi et al. 1999), but complex intra-day X-ray variability is present (e.g. Tanihata et al. 2001).

In 2013, Mkn~421 has undergone a prolonged state 
of high activity, with a peak around January 2013,  and revived episodes in April 2013.  Many instruments detected this high state: {\it Swift}, {\it Fermi}, 
{\it NuSTAR}, {\it MAXI}, ground-based optical, and TeV (Balokovi\'{c}  et al. 2013; Cortina, \& Holder 2013; Paneque et al. 2013; Negoro et al. 2013; Semkov et 
al. 2013; Krimm et al. 2013), and radio/mm wavelengths (Hovatta et al. 2013).  We activated our programme for {\it INTEGRAL}  follow-up of blazars in outburst, and started observing on 16 April 2013  with the IBIS/ISGRI, JEM-X, and OMC instruments.
The results of this campaign are presented here along with those of the simultaneous {\it Fermi}-LAT observations.


\begin{figure*}
   \centering
   \includegraphics[scale=0.5]{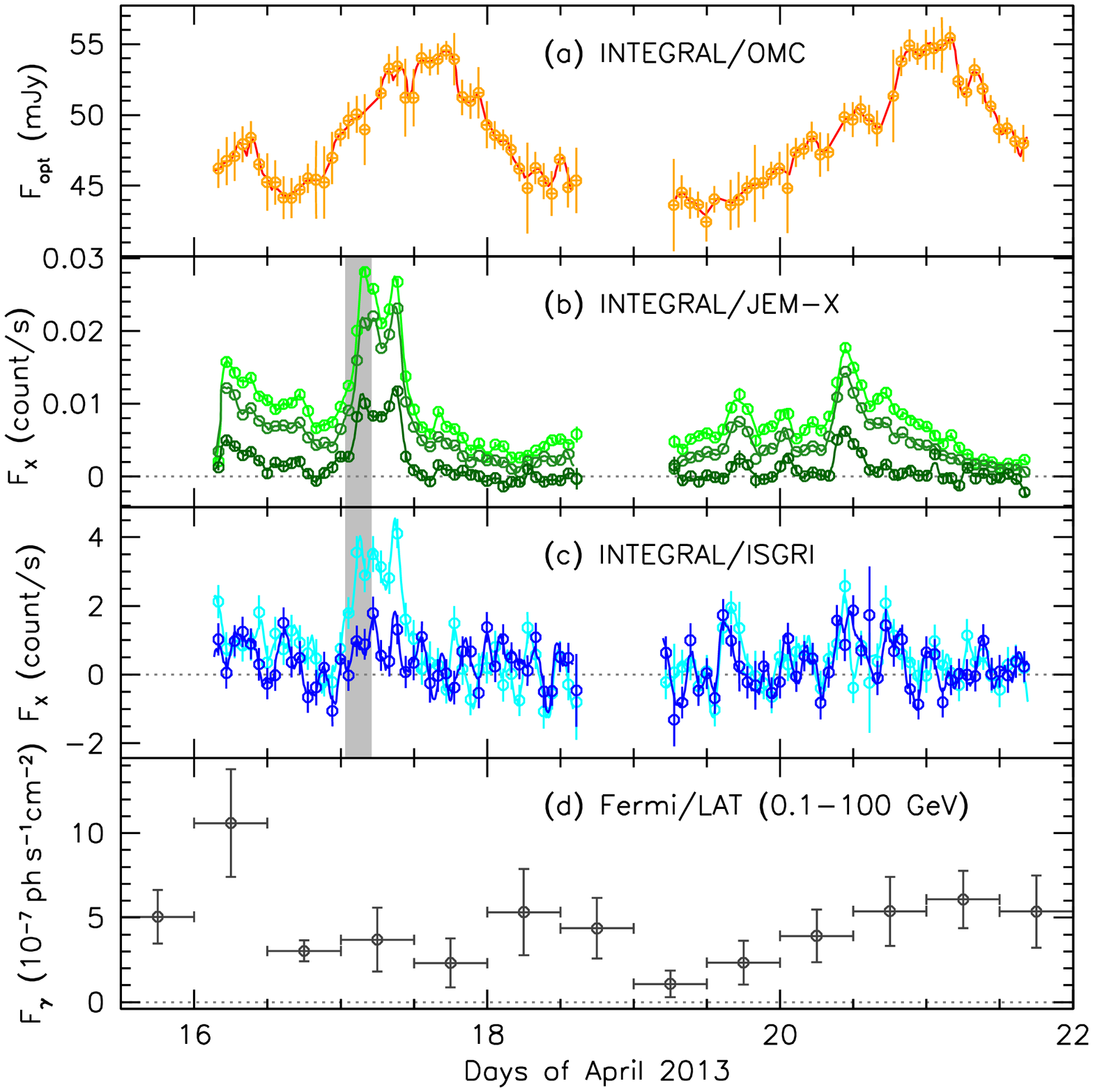}
      \caption{Light curves  of Mkn~421 in April 2013: 
      (a) OMC dereddened ($E_{B-V} = 0.014$) photometry, corrected for the host galaxy and for a companion galaxy that is located in the OMC field of view; 
      (b) JEM-X count rates (resulting from the coaddition of the signal measured by the 2 detectors),   in the bands 3.04-5.52 keV (light green),  5.52-10.24 keV (green), and 10.24-25.88 keV (dark green);
      (c) IBIS/ISGRI count rates at 20-40 keV (light blue) and 40-100 keV (dark blue);
      (d) {\it Fermi}-LAT fluxes binned with 12-hr time-resolution. 
                        As {\it INTEGRAL}  provides quasi-continuous datasets, we display them both as data points with a binning of 4.8 ks and as a curve representing a weighted smoothing of the data (see Sect. 3 for details).
            The shaded grey area in panels (b) and (c) represents the interval chosen for the cross-correlation analysis (see Fig.~\ref{FigDCF}).
}
         \label{FigLC}
\end{figure*}


\begin{sidewaystable*}
\begin{minipage}[t][180mm]{\textwidth}
\caption{Simultaneous JEM-X and IBIS/ISGRI  spectral fits$^a$ in the 3.5-60 keV range with a single (PL)  and broken (BPL) power law, and log-parabolic$^b$ (LP) models (uncertainties are 1-$\sigma$).}
\label{table:fitparamspl}
\centering
\begin{tabular}{cccccccccccccc}
\hline\hline             
UT & $\Gamma^c_{PL}$ & norm$_{PL}^d$ & $\chi^2_\nu$(d.o.f.) & $\Gamma_{BPL,1}$ & $\Gamma_{BPL,2}$ & $E_{break}^e$ & norm$_{BPL}^d$ & $\chi^2_\nu$(d.o.f.)  & $a$ & $b$ & norm$^f_{LP}$ & $\chi^2_\nu$(d.o.f.) & $F_{LP}^g$ \\
\hline

2013 Apr 16.13-16.52$^h$ & $2.58 \pm 0.04$ & $0.56 \pm 0.04$ & 1.69(32) & $2.2 \pm 0.1$ & $2.85 \pm 0.09$ & $6.5 \pm 0.9$ & $0.29 \pm 0.06$ &  0.95(30) & $2.67 \pm 0.05$ & $0.6 \pm 0.1$ & $1.60 \pm 0.05$ & 1.01(31) & $5.8 \pm 0.5$ \\

2013 Apr 16.93-17.1 & $2.58 \pm 0.06$ & $0.46 \pm 0.04$ & 0.87(32) & $2.43 \pm 0.09$ & $3.0 \pm 0.3$ & $10 \pm 3$  & $0.37 \pm 0.06$ & 0.72(30) & $2.66 \pm 0.08$ & $0.4 \pm 0.2$ & $1.28 \pm 0.06$ & 0.72(31) & $4.8 \pm 0.7$ \\

2013 Apr 17.1-17.42$^h$  & $2.49 \pm 0.02$ & $0.95 \pm 0.04$ & 2.64(44) & $2.21 \pm 0.05$ &  $2.82 \pm 0.07$ & $10 \pm 1$  &  $0.60 \pm 0.05$ & 1.27(42) & $2.49 \pm 0.02$ & $0.55 \pm 0.07$ & $3.47 \pm 0.07$ & 1.14(43) & $12.6 \pm 0.6$ \\    

2013 Apr 17.42-17.65 & $2.64 \pm 0.07$ &  $0.41 \pm 0.05$ & 1.40(32) & $1.9 \pm 0.4$ & $3.0 \pm 0.2$ & $5.4 \pm 0.9$  &  $0.13 \pm 0.08$ & 1.08(30) & $2.87 \pm 0.15$ & $0.9 \pm 0.3$ & $1.02 \pm 0.06$ & 1.08(31) & $3.6 \pm 0.7$ \\    

2013 Apr 20.21-20.48 & $2.65 \pm 0.06$ & $0.45 \pm 0.05$ & 0.77(32) &  $2.5 \pm 0.1$ & $2.9 \pm 0.2$ & $8    \pm 3$ & $0.35 \pm 0.07$ & 0.68(30) & $2.74 \pm 0.09$ & $0.4 \pm 0.2$ & $1.05 \pm 0.05$ & 0.67(31) & $4.0 \pm 0.6$ \\    
    
2013 Apr 20.48-20.78 & $2.65 \pm 0.05$ & $0.55 \pm 0.04$ & 1.76(32) & $2.2 \pm 0.2$  & $3.0 \pm 0.1$  & $6.6 \pm 0.9$  & $0.29 \pm 0.07$ & 1.25(30) & $2.9 \pm 0.1$ & $0.8 \pm 0.2$ & $1.29 \pm 0.05$ & 1.28(31) & $4.7  \pm 0.6$ \\    

quiescence$^{h,i}$  & $2.82 \pm 0.06$ & $0.24 \pm 0.02$ & 1.85(32) & $2.3 \pm 0.2$  & $3.3 \pm 0.2$ & $6.0  \pm 0.8$  & $0.12 \pm 0.03$ & 1.38(30) & $3.3 \pm 0.2$ & $1.1 \pm 0.3$ & $0.37 \pm 0.02$ & 1.46(31) & $1.4 \pm 0.3$ \\    

average & $2.65 \pm 0.02$ & $0.35 \pm 0.01$ & 3.1(32) & $2.3 \pm 0.1$ & $2.85 \pm 0.05$ & $6.5 \pm 0.6$ & $0.20 \pm 0.03$ &  1.71(30) & $2.70 \pm 0.03$ & $0.45 \pm 0.07$ & $0.85 \pm 0.01$ & 1.86(31) & $3.2 \pm 0.1$ \\

\hline 
\noalign{\smallskip}
\multicolumn{13}{l}{$^a$  The fit errors include a 3\% systematic uncertainty.} \\
\multicolumn{13}{l}{$^b$  In the log-parabolic model, $f(E) \propto (E/E_1)^{-(a + b~log(E/E_1))}$, the pivot energy $E_1$ was frozen to 10 keV.} \\
\multicolumn{13}{l}{$^c$ Photon index: $f(E) \propto E^{-\Gamma}$.}  \\
\multicolumn{13}{l}{$^d$ Normalization at 1 keV, in photons s$^{-1}$ cm$^{-2}$ keV $^{-1}$.}  \\
\multicolumn{13}{l}{$^e$ In keV.}  \\
\multicolumn{13}{l}{$^f$ Normalization at 10 keV, in $10^{-3}$ photons s$^{-1}$ cm$^{-2}$ keV $^{-1}$.}  \\
\multicolumn{13}{l}{$^g$  Flux intensity in 3.5-60 keV from the log-parabola model, in $10^{-10 }$ erg s$^{-1}$ cm$^{-2}$.}  \\
\multicolumn{13}{l}{$^h$  This epoch is represented  in Figure~\ref{FigSED}.}  \\
\multicolumn{13}{l}{$^i$  The spectral signal was integrated over the two intervals 2013 April 17.77 to 19.58 and April 21.04 to 21.7 UT.}  \\

\end{tabular}
\vfill
\end{minipage}
\end{sidewaystable*}


\section{Observations and results}

\subsection{{\it INTEGRAL}}

Mkn~421 was observed as a Target of Opportunity by {\it INTEGRAL} (Winkler et al. 2003) in the periods 2013 April 
16.13-18.58 UT (Revolution 1283) and 19.20-21.68 UT (Revolution 1284) for 200 ks each time. The effective net exposure  time  was 270.9 ks for IBIS/ISGRI (Ubertini et al. 
2003; Lebrun et al. 2003) and  345.7 ks  for JEM-X (Lund et al. 2003).  We observed in hexagonal dithering mode, so that the source was always in the JEM-X field of view.  The   screening, reduction, and analysis of the {\it INTEGRAL} data have been performed using the {\it INTEGRAL} Offline 
Scientific Analysis (OSA) V. 10.0, publicly available through the {\it INTEGRAL} Science Data Centre (ISDC, 
Courvoisier et al. 2003). The algorithms implemented in the software are described in Goldwurm et al. (2003) 
for IBIS and Westergaard et al. (2003) for JEM-X.   

\subsubsection{OMC}

The OMC  (Mas-Hesse et al. 2003)   data  were acquired with a standard 
V-band Johnson filter and extracted with default settings, using a 3x3 pixels binning, which is 
appropriate for point-like sources.  Individual measurements that were flagged as problematic were  disregarded. The target is well detected.   From the OMC  data-points  we  subtracted both  the contribution of the host galaxy and that of a companion galaxy located at about 14 arc-sec north-east of the nucleus (see Ulrich 1978;  Gorham et al. 2000; Nilsson et al. 2007).  For the former we adopted $R = 13.29 \pm 0.02$ from {\it Hubble Space Telescope} imaging and a colour $V - R = 0.63$ (Urry et al. 2000).  This magnitude was obtained by integrating the galaxy radial profile to infinity, which is appropriate, considering that the adopted OMC photometric aperture corresponds to a diameter of 50 arc-sec, which is equivalent  to infinity for practical purposes (the host of Mkn~421 has a half-light radius of only  a few arc-sec).   For the companion galaxy we used the SDSS photometry reported in the NED\footnote{ned.ipac.caltech.edu} database (this source is identified as RXJ1104.4+3812:BEV[98]014), converted to the Johnson V band. Our total estimated flux of this satellite  galaxy in the V-band is $1.72 \pm 0.02$ mJy.   Finally, we have applied a Galactic extinction correction using $E_{B-V}$ = 0.014 (Schlafly \& Finkbeiner 2011) and the curve of Cardelli et al.  (1989). 

The corrected OMC fluxes  are reported in Figure~\ref{FigLC}a.  
Two  flares, of which only the first is fully resolved,   are detected.  The    first outburst is symmetrical, with rising   and decay times (defined as the intervals during which the flux increases or decreases from quiescence to peak, respectively) of 1 day each in the observer frame.  Taking the uncertainties into account, the variation amplitude from quiescence to maximum flux is 20\%.  The second outburst shows an increase of  similar amplitude in 1.5 days, while the decay is not fully sampled.  Our average flux  ($48.5 \pm 0.4$ mJy)  is approximately consistent with the photometry  obtained just prior to our monitoring (Semkov et al. 2013), and it is  about a factor of 2 larger than measured in June 2006 ($\sim$25 mJy) by Lichti et al. (2008), when it is taken into account  that these authors adopt a somewhat dimmer host galaxy contribution, and do not correct for the presence of the host galaxy companion.

\subsubsection{JEM-X and IBIS}

The source is detected with JEM-X  in most individual pointings with a significance of 10-$\sigma$.  
The light curve is reported in Figure~\ref{FigLC}b  in three different bands.

IBIS/ISGRI detected the source  significantly in individual science windows ($\sim$2000 s) only on some occasions, notably  during  maximum flux on April 17, when it reached 8-$\sigma$ significance in a single science window.   The final mosaic yields  an average flux in the 20-100 keV range of  $1.10 \pm 0.05$ counts s$^{-1}$, corresponding to ($4.4 \pm 0.2$) mCrab.  The IBIS/ISGRI light curves in the distinct ranges 20-40 keV and 40-100 keV are reported in Figure~\ref{FigLC}c.   No detection was obtained with IBIS/PICsIT.

As in the optical band, the X-ray observations display two main outbursts detected up to $\sim$40 keV, with much higher amplitude  (a factor of $\sim$4 in 5-10 keV in the first outburst and a factor of $\sim$3 in the second outburst with respect to quiescence) and shorter timescale: while the first outburst appears approximately symmetrical with rising and decay times of $\sim$0.5 days each,  in the second outburst the rise time is less than 0.5 days while the decay lasts about 1 day.
The 40-100 keV light curve is affected by large statistical errors, so that assessing variability is difficult.

Our JEM-X measurement of the X-ray flux  is comparable to that of June 2006  reported by Lichti et al. (2008), while our IBIS/ISGRI measurement is about a factor of 2 lower.


\begin{figure*}

\begin{tabular}{ccc}

\begin{minipage}{0.28\textwidth}
  \includegraphics[width=1.4\textwidth]{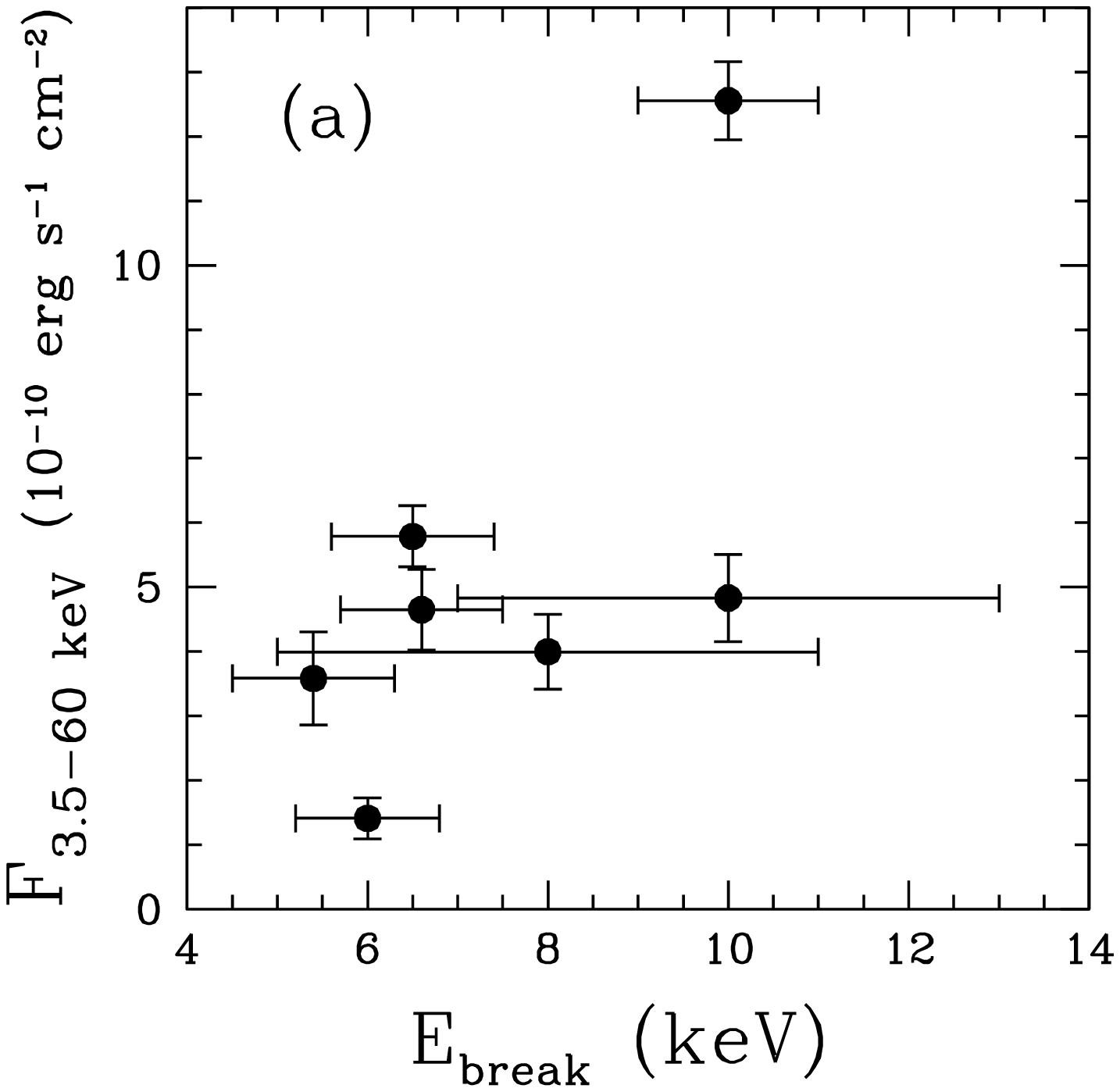}
\end{minipage} 
&
\begin{minipage}{0.28\textwidth}
  \includegraphics[width=1.4\textwidth]{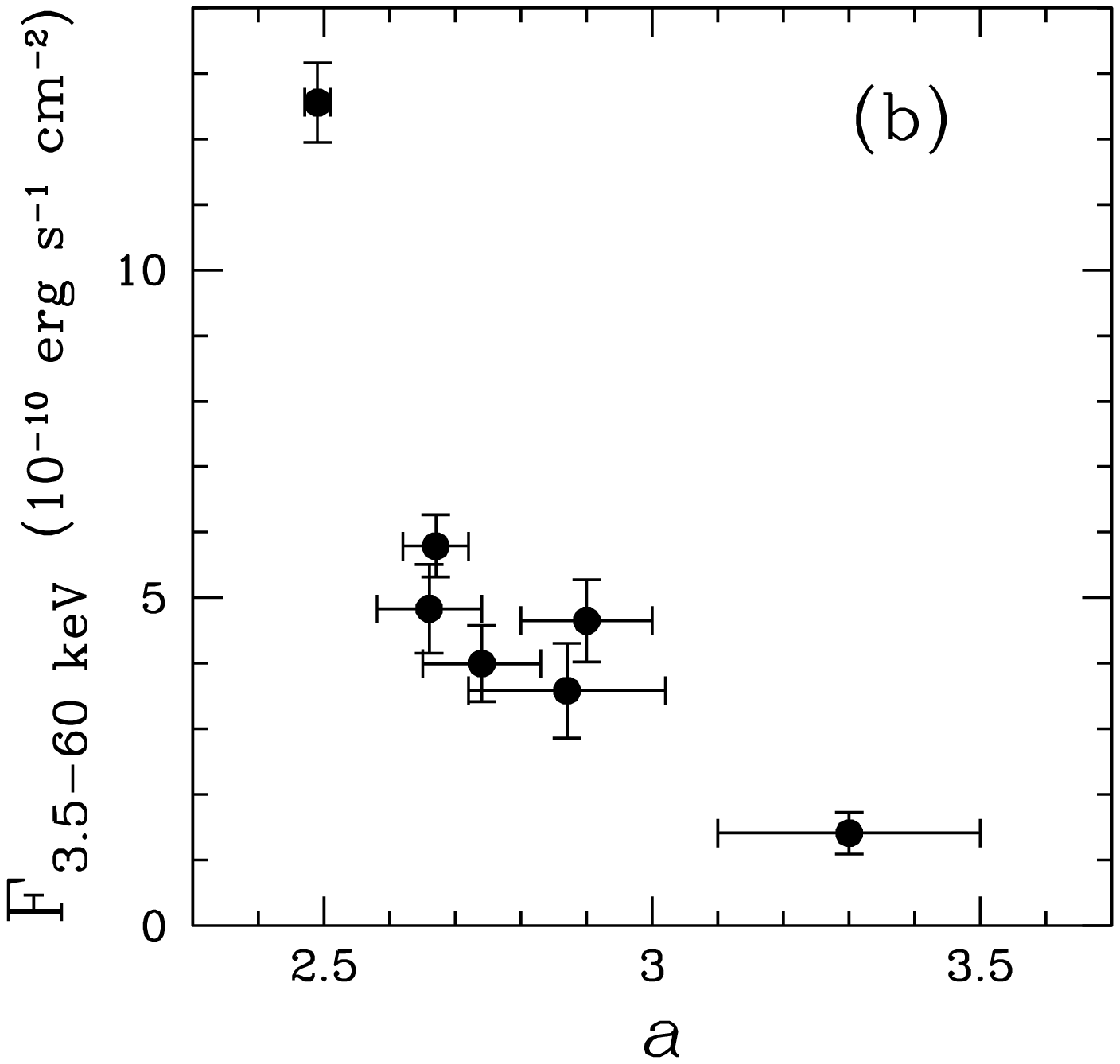}
\end{minipage}
&
\begin{minipage}{0.28\textwidth}
  \includegraphics[width=1.4\textwidth]{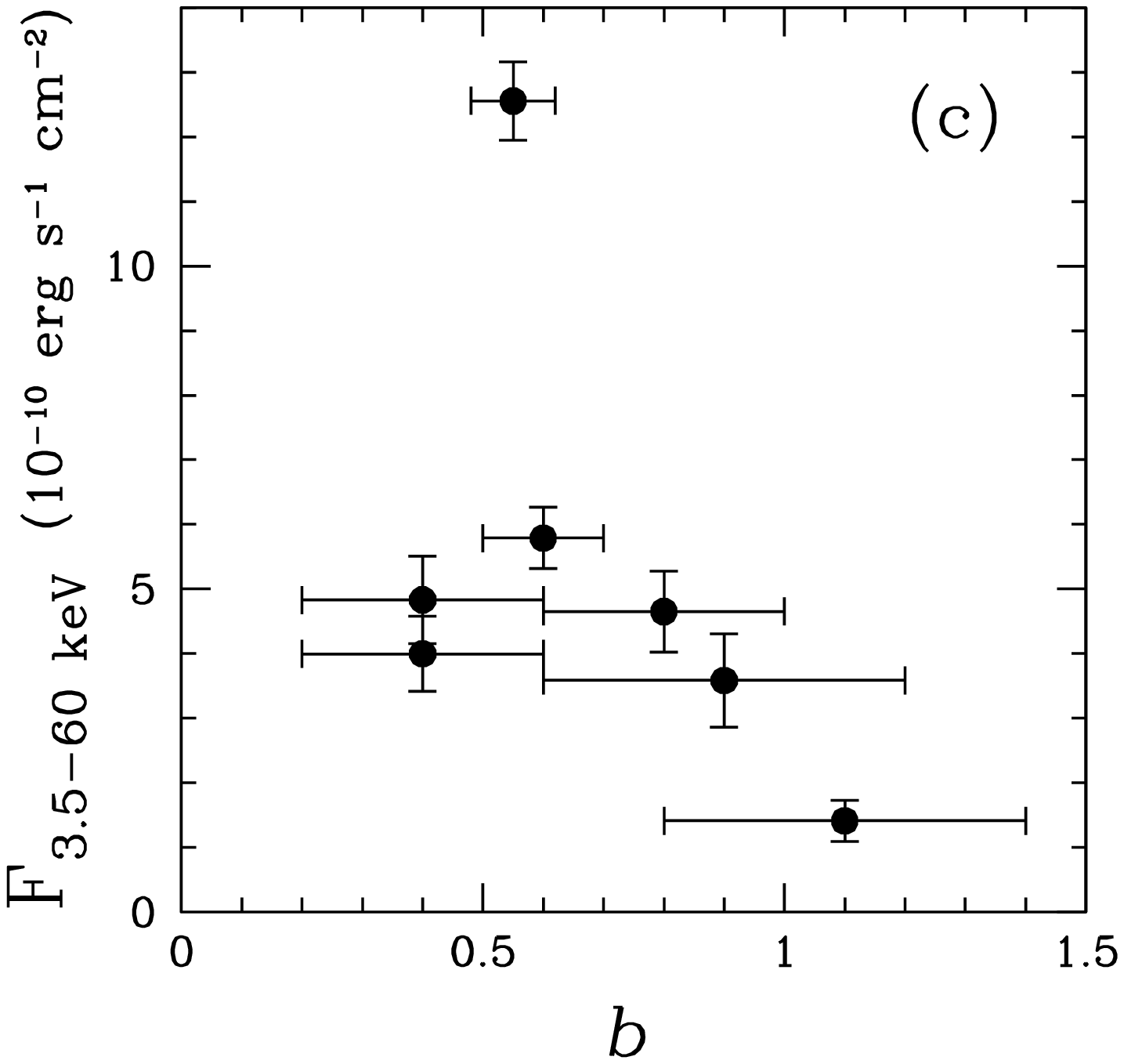}
\end{minipage}

\end{tabular}

\caption{Flux integrated in the joint JEM-X and IBIS/ISGRI range (3.5-60 keV), as derived from the log-parabola spectral  fits,  vs (a) break energy, as derived from the broken power law fits of time-resolved spectra; (b) $a$ index from log-parabola fits; (c) $b$ index from log-parabola fits.   }
         \label{FigIBISfluxvsspec}

\end{figure*}

\subsubsection{X-ray  spectra}

The overall JEM-X and IBIS/ISGRI spectra were extracted, combined and, after convolution with the  most recent matrices  available in OSA V.  10.0, fitted in the energy range 3.5--60 keV with a single  power law, 
a broken power law, and with a log-parabola of the form $f(E) \propto [E/(10 keV)]^{-(a + b~log[E/(10 keV)]}$ (Massaro et al. 2004a, 2006; Perlman et al. 2005).   A fixed inter-calibration constant of unity between JEM-X and IBIS/ISGRI  was assumed (Jourdain et al. 2008).  Photoelectric absorption caused by the neutral hydrogen along this line of sight ($N_{\rm H} = 1.43 \times 10^{20}$ cm$^{-2}$, Elvis et al. 1989) is negligible at these frequencies, and therefore not corrected for.  The fit results are reported in the last row of Table~\ref{table:fitparamspl}. 

In order to study spectral variability in different states of the source, we defined a series of time intervals for which we extracted JEM-X 1 \& 2 spectra and IBIS/ISGRI spectra. We chose to investigate separately the rising and the decaying phases of the two main flares, i.e. the periods April 16.93--17.1 and April 17.42--17.65 for the first flare, and the periods April 20.21--20.48 and April 20.48--20.78 for the second flare. 
In addition, we selected the time period April 16.13--16.52, which is  simultaneous to the phase of highest GeV emission,  the extended period of high flux during the first flare from April 17.1 to 17.42, and a quiescent state obtained by coadding the signal during the periods April 17.77 to 19.58 and April 21.04 to 21.7, to increase   the signal-to-noise ratio.

We then fitted the  joint JEM-X and IBIS/ISGRI spectra (3.5-60 keV)  with  a single power law,  a broken power law, and a log-parabola model, using the same assumptions for JEM-X vs IBIS/ISGRI inter-calibration and Galactic absorption as adopted for the average spectrum. The results are reported in Table~\ref{table:fitparamspl} and in 
Figure~\ref{FigIBISfluxvsspec}.  For the average and all time-resolved joint JEM-X and IBIS/ISGRI spectra the broken power law  and the log-parabola yield better fits than the single power law, and the log-parabola is preferable over the broken power law because it is marginally better constrained (3 vs 4 free parameters).    While the break energy parameter of the broken power law fits and the {\it a} parameter of the log-parabola model both vary significantly (the constancy probability is less than 0.02), the  log-parabola {\it b} parameter only varies marginally (constancy probability of 0.35).
Accordingly, the correlation of the break energy and  {\it a} parameter  with flux is more significant than that of the {\it b} parameter with flux:
the weighted linear correlation coefficient between the break energy and the integrated flux is $r = 0.73$, with a probability of 0.062 of chance correlation; the weighted linear coefficients between the  log-parabola  indices {\it a} and  {\it b}  and the flux are $r = -0.85$ (probability of 0.015) and $r = -0.66$ (probability of 0.11), respectively (Bevington \& Robinson 2003).


\begin{figure*}
\centering
\includegraphics[scale=0.9]{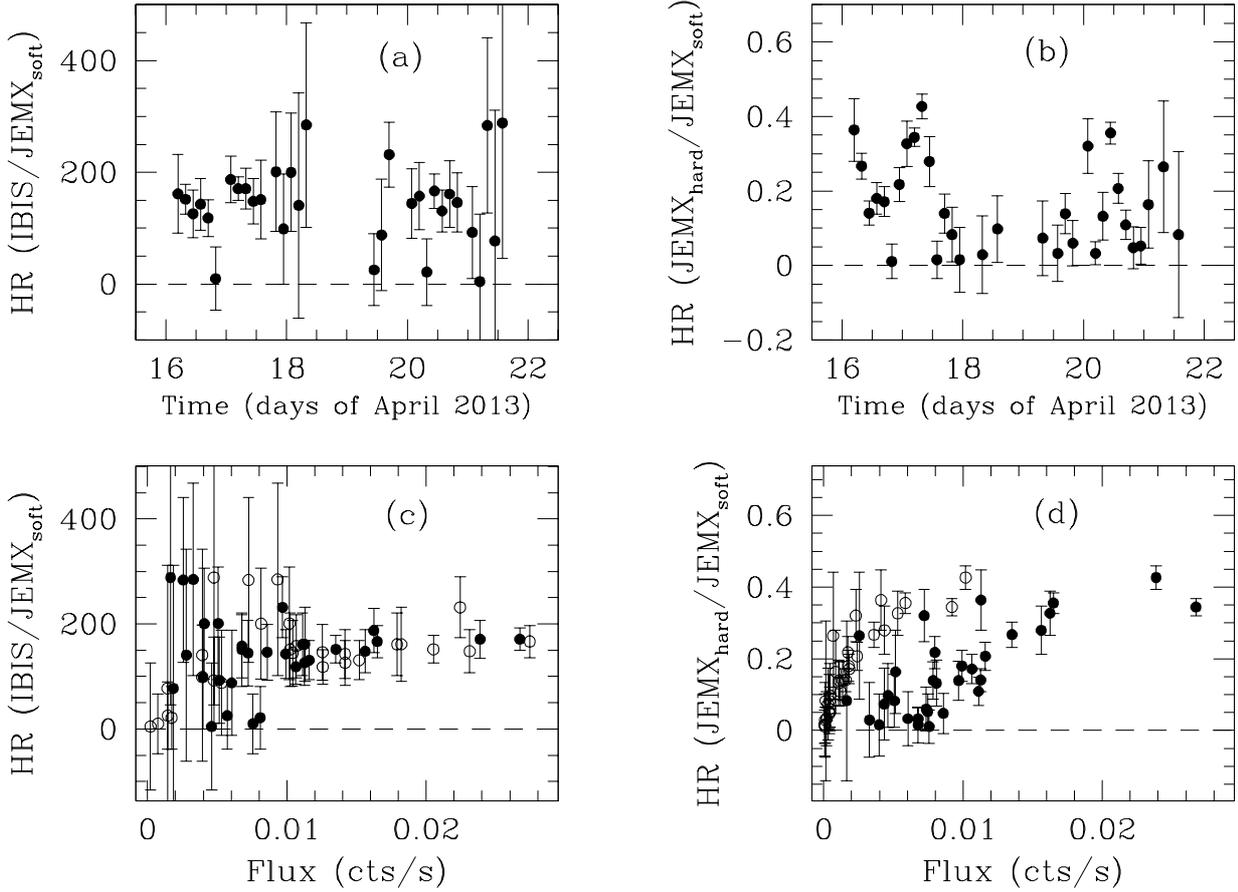}
\caption{
(a) Ratio between IBIS/ISGRI  (20-100 keV)  flux and JEM-X (3.04-5.52 keV) flux as a function of time;   
(b) ratio between JEM-X 10.24-25.88 keV flux and 3.04-5.52 keV flux as a function of time;
(c)  ratio between 20-100 keV flux vs 3.04-5.52 keV flux as a function of  3.04-5.52 keV  flux (filled circles) and  20-100 keV flux reduced by a factor of 100 (open circles);  
(d) ratio between 10.24-25.88 keV flux and 3.04-5.52 keV flux as a function of 3.04-5.52 keV (filled circles) and 10.24-25.88 keV flux multiplied by 2 (open circles).  In all cases, the adopted time binning is 3 hours.}
         \label{Fighratios}

\end{figure*}

\subsubsection{Hardness ratios}

As done in Lichti et al. (2008), we have computed the hardness ratios in the energy  bands covered by IBIS/ISGRI and JEM-X.  Specifically, we have first computed the ratio between the simultaneous fluxes in the 20-40 keV  and 40-100 keV ranges.   Unlike in the IBIS/ISGRI observations of June 2006, when the gamma-ray flux was higher than presently found (Lichti et al. 2008),  this ratio does not vary significantly within the large errors, nor is there any significant correlation of the hardness ratio with either the softer or the harder IBIS/ISGRI band flux.  Then we evaluated the ratio between the simultaneous fluxes in the whole IBIS/ISGRI range (20-100 keV) and in the softest JEM-X band (3.04-5.52 keV).  The variability behaviour of this ratio traces that of the  light curves (Fig.~\ref{FigLC}), i.e. it indicates spectral hardening accompanying flux brightening.  Although the result has a limited significance (see 
Fig.~\ref{Fighratios}a), it is confirmed by the correlation of hardness ratio with both the gamma- and X-ray flux 
(Fig.~\ref{Fighratios}c), and it is consistent with the findings of spectral analysis (see Fig.~\ref{FigIBISfluxvsspec}).
Finally we have computed the ratios of the simultaneous fluxes in the JEM-X ranges 3.04-5.52 keV and 10.24-25.88 keV.  
These follow a  behaviour similar to that of the IBIS/ISGRI vs JEM-X hardness ratios:  the  ratio between these X-ray bands varies in a similar way as the JEM-X light curves  (Fig.~\ref{Fighratios}b)  and it increases with flux, both when the 10.24-25.88 keV and the   3.04-5.52 keV  fluxes are considered  (Fig.~\ref{Fighratios}d).  Again, this indicates correlated spectral hardening with source brightening, as often seen in Mkn~421 at these energies.  No different dependencies are seen between hardness ratio and flux during brightening vs dimming phases (hysteresis cycles), as previously reported by Takahashi et al. (1996) and Fossati et al. (2000b) in X-rays.


\begin{table*}
\caption{Statistical parameters of {\it Fermi}-LAT fluxes and spectral indices  correlation$^a$} 
\label{table:latstatparams}
\centering                       
\begin{tabular}{ccccccc}      
\hline    
\hline
Test  & $\chi^2_{12h}$  &  $\nu_{12h}$  &    P$^b_{12h}$  &  $\chi^2_{24h}$  &  $\nu_{24h}$  &    
P$^b_{24h}$   \\
\hline                 
Constant photon index           &  9.93 (4.37)$^c$  & 10  (9)  & 0.50 (0.89)   &  6.16  &  6  &  0.47 \\
Constant flux                          &  17.3 (13.0)         &  10 (9)  &  0.065 (0.18) & 21.9  &  6  &  $< 10^{-4}$ \\
Weighted least squares linear fit   &   7.0  (7.0)    &  9 (8)    &  0.68  (0.59)      &  4.50    &  5  & 0.54 \\
Weighted correlation coefficient    &  0.77 (0.68)  &  9 (8)    &  0.0055 (0.03)   &  0.89   &  5 &  0.0072 \\
\hline  
\noalign{\smallskip}
\multicolumn{7}{l}{$^a$  The sub-index in the $\chi^2$ value, degrees of freedom $\nu$ and probability P indicates  the time}\\
\multicolumn{7}{l}{ ~~~ binning of the data:  12 or 24 hours.}\\
\multicolumn{7}{l}{$^b$  This indicates the probability of the null hypothesis for the tests of constancy and
linear}\\
\multicolumn{7}{l}{ ~~~  trend and of chance correlation for the correlation coefficient.}\\
\multicolumn{7}{l}{$^c$  The quantities in parentheses for the 12-hr binning parameters refer to the case where the}\\
\multicolumn{7}{l}{ ~~~  point of steepest spectrum and  highest flux was removed.}
\end{tabular}                               
\end{table*}


\subsection{{\it Fermi}-LAT}

Data of Mkn~421 covering the period 15-22 April 2013  were downloaded from the LAT online public archive\footnote{http://fermi.gsfc.nasa.gov/cgi-bin/ssc/LAT/LATDataQuery.cgi}  and analysed with standard methods\footnote{http://fermi.gsfc.nasa.gov/ssc/data/analysis/scitools/}.
In particular, we used a  zenith angle of 100 degrees and a rocking angle of 52 degrees.
The adopted software is  LAT Science Tools v. 9.27.1,  with the Instrument Response Function  P7SOURCE\_V6 and  corresponding files for the Galactic diffuse and isotropic background.  The extraction region has a radius of interest  of  10 degrees and is centred on the radio position of Mkn~421.   
The sources of the  2FGL Catalog  present within this radius and Mkn~421 itself were modelled with single power laws  by leaving the parameters free.   We have verified that by adopting a somewhat bigger radius, 15 degrees, our results do not change.  In order to verify the robustness of the source detection we used the Test Statistics (TS) method (Mattox et al. 1996), taking a threshold of TS = 25 (equivalent to about 5 $\sigma$).  
The light curve  is reported in Figure~\ref{FigLC}d.

\begin{figure}
   \centering
   \includegraphics[scale=0.4]{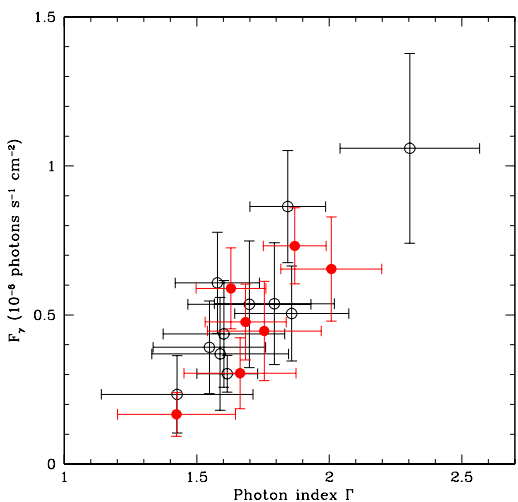}
      \caption{{\it Fermi}-LAT flux (0.1-100 GeV) vs photon index $\Gamma$ during the campaign.  The data are averaged over intervals of 12 hours (black open circles) and 24 hours (red filled circles).   Only measurements for which $TS > 25$ were retained.}
         \label{FigLAT}
\end{figure}

The LAT state is similar and marginally brighter than detected by Abdo et al. (2011) and a factor of 3 brighter than reported in the 2FGL catalog (Nolan et al. 2012), after reducing all fluxes to the 0.1-100 GeV range. 
We extracted the LAT spectra over time intervals of 12  and 24 hours and fitted them to single power laws ($f(E) \propto E^{-\Gamma}$) in the range 0.1-100 GeV.   
A correlation between flux and photon index is seen in the sense of spectral hardening for flux dimming 
(Fig.~\ref{FigLAT}), which is similar to 
previous behaviour seen in Mkn~421 in the period August 2008 - March 2010, and in another blazar of the HBL class,  PKS~2155-304 (Foschini et al. 2010).  We have tested this correlation quantitatively (see e.g. Albert et al. 2007b; Acciari et al. 2009; Aharonian et al. 2009; Anderhub et al. 2009).  We performed a test of constancy both on photon indices and fluxes, an error-weighted least squares linear fit and a weighted linear Pearson correlation coefficient, and computed the associated probabilities according to Bevington \& Robinson (2003).  All results are reported in 
Table~\ref{table:latstatparams}.   The 12-hr binned photon index is consistent with a constant behaviour,  while the flux binned with the same time resolution  varies somewhat more significantly.  Both the weighted linear fit and the correlation coefficient suggest however a decent correlation, both when the whole dataset is used and when the point of highest flux and steepest spectrum is ignored (see Fig.~\ref{FigLAT}).  This suggests that the photon index and flux, albeit  modestly variable, are correlated.  When the 24-hr binned fluxes and indices are considered, this conclusion is confirmed.
We note that for that same period, Abdo et al. (2011) do not report any  correlation between flux and photon index  for Mkn~421, likely because it cannot be seen with integration times of one week.


\section{Variability and timing analysis}

Following Lichti et al. (2008), we evaluated the variability of our  multiwavelength light curves using a fractional rms variability amplitude as defined in Fossati et al. (2000a)  and Vaughan et al. (2003).  As in Lichti et al. (2008), we find increasing variability at increasing energies, although this trend is reverted at the LAT energies, where the fractional variability is lower than that computed for IBIS/ISGRI  (see Table~\ref{table:fracrmsvar}), albeit marginally significant.   This is partly due to the relatively large errors of the LAT measurements, but it also matches the fact that the GeV photons are produced by inverse Compton scattering off the electrons that are responsible for the less variable optical spectrum.  With respect to Lichti et al. (2008), our variability indices are all larger, indicating larger variability amplitudes in general and a higher level of inter-day and intra-day activity.  

We also attempted to cross-correlate the JEM-X and IBIS/ISGRI light curves in search of time lags.  
We extracted the light curves in various energy ranges in  time bins of 300 s for JEM-X and 600 s for IBIS/ISGRI and
smoothed these oversampled light curves  in a time window of $\sim$6 ks with a modified boxcar smooth (T\"urler, in prep.).  The averaging inside the time bins is done by taking into account both the measurement errors and sampling times. More weight is given to points with lower uncertainties and  closer to the centre of the time window, making the smoothing robust for light curves with an irregular sampling and unequal errors.
We used the Interpolated Cross-Correlation Function (ICCF, Gaskell \& Peterson 1987), which is appropriate for curves with no big data gaps,   with the improvements introduced by White \& Peterson (1994).  The time lag corresponds to the maximum of the correlation curve.     

\begin{figure}
   \centering
   \includegraphics[scale=0.3]{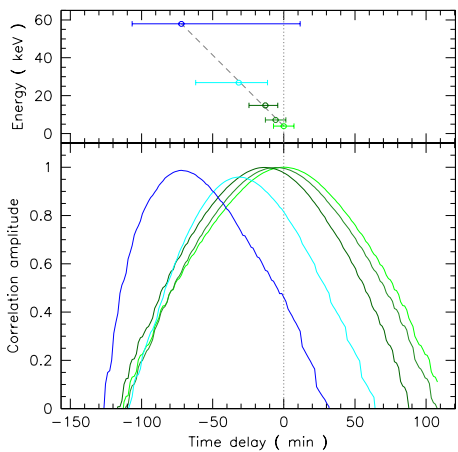}
      \caption{Cross-correlation function of the {\it INTEGRAL} JEM-X (the outputs of the 2 detectors have been coadded) and IBIS/ISGRI  light curves during the onset and first peak of the main outburst  (April 17.03-17.21 UT).   Bottom panel: correlation curves of JEM-X data in the softest range (3.04-5.52 keV) with itself and with the light curves at higher energies, identified by the same colours as in Figure~\ref{FigLC}.  Top panel: corresponding correlation time lags as a function of  energy (same colour-coding as in bottom panel) with 1-$\sigma$ uncertainties based on the distribution of lags obtained by perturbing the original light curves (see text).    The energy was evaluated by calculating the average photon energy in the spectral band assuming a photon index of $\Gamma=2.8$, as obtained by a single power law fit to the combined JEM-X and IBIS/ISGRI overall spectra.   The dashed line in the upper panel simply connects the first and last points to visualise the trend.}
         \label{FigDCF}
\end{figure}

In Figure~\ref{FigDCF}  (bottom panel)  we show the correlation curves of the JEM-X 3.04-5.52 keV light curve with respect to itself (auto-correlation), and with respect to the other bands during  the rising phase of the first flare,  i.e. when only data in the time window  April 17.03-17.21 UT are considered.   
In the top panel of the figure are reported  the corresponding time lags vs the  energy of the band centroid.   A negative time lag corresponds to harder photons leading the softer ones.   The 1-$\sigma$ errors  are  obtained by calculating the lag on 1000 perturbed light curves and following the  flux-randomization/random subset-selection method described in Peterson et al. (1998).   These perturbed light curves were then smoothed and cross-correlated with the ICCF method. 
We estimated the 1-$\sigma$ statistical uncertainties by taking the boundaries of the region comprising the central 68\% of the distribution of their time  lags  (i.e. $\pm$ 34\,\% from the median).
The  higher energies light curves lead the softest one by a linearly increasing time lag (a weighted linear least squares fit of the 5 time lags vs their corresponding energies yields $\chi^2 = 0.118$ with a probability of 99\%), with a maximum  of 72 minutes between the 40-100 keV and the 3.04-5.52 keV signal.  However, this trend is formally not significant because the time lags deviations from a constant lag of -7 minutes correspond, when error-weighted, to a  $\chi^2 = 3.15$ (assuming the upper errors on the time lags, that are on average worse),  which yields  a constancy probability of  59\% for 4 degrees of freedom.

We explored other time windows around the first and second flares and did not find more compelling evidence for time delays. In particular, the  flare of April 20 suggests a similar trend for JEM-X, but the IBIS/ISGRI signal is not significant, whereas the peak and early decay of the April 17 flare is found to be quasi-simultaneous in all X-ray bands.
Our  time lags are comparable in absolute value to those found for Mkn~421 in the {\it INTEGRAL} observations of  June 2006 (Lichti et al. 2008), and to those reported by Sembay et al. (2002), Brinkmann et al. (2003), Ravasio et al. (2004) for the X-rays only (see also Zhang et al. 2004).   They are also in line with the report of a much larger time lag, 10 days, of the optical vs X-ray photons (Gaur et al. 2012).  These lags can set important constraints on the cooling times of the relativistic particles responsible for the synchrotron radiation, and in turn on the magnetic field.  


The two maxima in the OMC light curve might be related to the two main outbursts seen in the X-rays. The overall delay is of $\sim$0.5 days, as confirmed by a cross-correlation analysis.
We did not try the correlation test on  the LAT light curve, because of the paucity of flux points.


\begin{table}
\caption{Fractional rms variability amplitude} 
\label{table:fracrmsvar}
\centering                       
\begin{tabular}{ccc}      
\hline\hline              
Instrument & Band   & $F_{var}^a$ \\
\hline                 
OMC            &  V-band                   &  $0.065 \pm 0.003$ \\  
JEM-X           &  3.04-5.52 keV        &  $0.68  \pm 0.01$\\     
JEM-X           &  5.52-10.24 keV      &  $0.85 \pm 0.01$ \\   
JEM-X           &  10.24-25.88 keV    &  $1.78 \pm 0.04$ \\   
IBIS/ISGRI   &  20-40 keV              &  $1.4 \pm 0.1 $ \\   
IBIS/ISGRI   &  40-100 keV            &  $1.5 \pm 0.3$  \\   
{\it Fermi}-LAT    &  0.1-100 GeV          &  $0.35  \pm 0.14 $   \\   
\hline  

\noalign{\smallskip}
\multicolumn{3}{l}{$^a$  See  Fossati et al. (2000a), Vaughan et al. (2003).}\\
\end{tabular}                               
\end{table}


\section{Discussion}

\begin{figure*}
   \centering
   \includegraphics[scale=0.5]{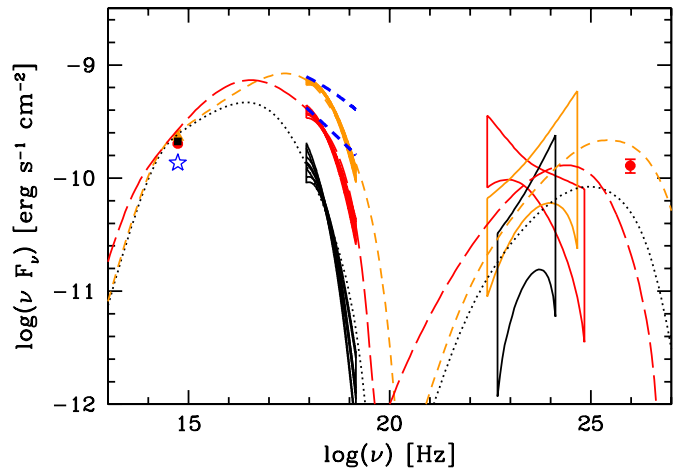}
      \caption{Spectral energy distributions of Mkn~421  at the average UT epochs of 16.1--16.5 (red circle) and 17.1--17.4 (orange triangle) April 2013 and during the quiescent  state (black square), that we relate here    to the minimum {\it Fermi}-LAT flux of April 19.0--19.5,  from simultaneous {\it INTEGRAL} IBIS/ISGRI, JEM-X and OMC, and {\it Fermi}-LAT data.  The optical data were corrected for Galactic absorption and for the contamination by galaxies in the field as described in the text.  The 1-$\sigma$ error contours of the  joint JEM-X and IBIS/ISGRI spectra and LAT spectra are  reported.  The TeV point (red) was taken on April 16.3 and is the last point of the VERITAS observation (Mukherjee 2013;   still preliminary).  The models (long dash:  April 16, short dash: April 17, dot:  April 19) include a synchrotron component at the lower energies, produced in a single emitting zone, and a synchrotron self-Compton scattering component at  higher energies (see model parameters in Table~\ref{table:sedmodelparams}).  For comparison, the JEM-X and IBIS/ISGRI quiescent and active state spectra of June 2006 (Table~7 in Lichti et al. 2008) are reported as thick dashed blue lines.  The average optical flux at the same epoch is shown as a  blue star.}
         \label{FigSED}
\end{figure*}

We observed and detected Mkn~421 with the {\it INTEGRAL} instruments IBIS/ISGRI, JEM-X and OMC during a high state that followed the detection of a powerful TeV energy outburst  (Cortina \& Holder 2013)   that was likely subsiding during our observation. The X-ray flux is among the highest previously recorded for this object   by {\it INTEGRAL} itself and other satellites (Malizia et al. 2000; Donnarumma et al. 2009; Abdo et al. 2011).    The optical state is also very bright, with few precedents  (Tosti et al. 1998).    While our  X-ray flux  is similar to that reported by Lichti et al. (2008) for the observations of June 2006,  our optical and gamma-ray measurements are a factor of 2 higher and lower, respectively, suggesting that the synchrotron spectrum pivoted around the X-ray frequencies during the 2006 and 2013 states.  This also results in a lower break energy than observed by Lichti et al. (2008).  This parameter never exceeds 10 keV  in our fits,  while the X-ray models of Lichti et al. (2008) can accommodate a break energy higher than $\sim$40 keV.  
As suggested by both our models and those presented by Lichti et al. (2008), the peak of the $\nu f_{\nu}$ synchrotron spectrum occurs at lower energies ($\sim$1 keV in our multiwavelength models, see Fig.~\ref{FigSED}). 

Besides a broken power law, we have attempted to fit the joint JEM-X and IBIS/ISGRI spectra  with a smoothly steepening power law, best described by a log-parabola model, as it is often seen in BL Lacs where the hard X-ray radiation is due to the synchrotron process (Maraschi et al. 1999; Tavecchio et al. 2001; Massaro et al. 2004b; Lichti et al. 2008; Giommi et al. 2012).    
The X-ray flux computed from the log-parabolic fits correlates decently both with  the break energy derived from the broken power law fits, and with the log-parabola {\it a} index (Fig.~\ref{FigIBISfluxvsspec}).   A correlation between flux and break energy and/or spectral indices is expected, because  injection of fresh and energetic particles  in the emitting region should cause both flux enhancement and spectral hardening.    This is seen, with various levels of significance, also in the behaviour of the JEM-X and IBIS/ISGRI hardness ratios  (Fig.~\ref{Fighratios}) and in the frequency-dependent variability (Table~\ref{table:fracrmsvar}).
 
We  found a weak correlation between LAT flux and photon index, in the sense of a softer spectrum accompanying a higher flux (Fig.~\ref{FigLAT}), which is opposite of what we observe in the {\it INTEGRAL} data.  This  can be explained if the effect is highly dependent on the observing band:  when peak energy of the inverse 
Compton component    is located within 
the LAT range the flux is higher and the spectrum is flat ($\Gamma \sim 2$), while when it moves  to energies higher than the LAT range it causes the  flux to decrease and the spectrum to harden ($\Gamma < 2$). Notably, this behaviour seems to be only weakly correlated with the X-ray variations   (Fig.~\ref{FigLC}), and is not observed on timescales longer than $\sim$1 day (Abdo et al. 2011). 
Whether systematics or instrumental effects are involved it is difficult to assess, because  the phenomenon may  trace complex physics, whose signal  washes out when mediated on long timescales.

Multiwavelength spectral energy distributions of Mkn~421 from optical to gamma rays were constructed   in three representative  states (Fig.~\ref{FigSED}): the first state refers to the initial phase of our {\it INTEGRAL} observation (April 16.13-16.52), when LAT measured the brightest flux and softest spectrum; the second state represents the first {\it INTEGRAL} flare (April 17.1-17.42), when the 3.5-60 keV flux was at its brightest and LAT measured an intermediate flux and spectrum; the third state describes the dimmest and hardest LAT measurement of April 19.0-19.5, associated with the quiescent {\it INTEGRAL} state (indicated with  ``quiescence" in Table~\ref{table:fitparamspl}), that we obtained by averaging the flux over two intervals where the 3.5-60 keV flux was lowest.    The TeV  flux  had subsided to quiescence 
by the time our {\it INTEGRAL} observation started,  thus we report here the latest flux ($1.75 \pm 0.25$ Crab units at $E  >  400$ GeV)  of the  campaign of the VERITAS  Cherenkov telescope (Mukherjee 2013), which, recorded on April 16.3 UT,  is quasi-simultaneous with the start of our observation.  We caution the reader that this measurement is still preliminary.  In Figure~\ref{FigSED} are reported  for comparison  also the observations of June 2006 (Lichti et al. 2008), that show the similarity of the X-ray flux and the difference in  optical and gamma-ray state.

We modelled the spectral energy distributions with a single-zone emitting model with a synchrotron component at lower energies, and self-Compton scattering at the higher energies (e.g. Tavecchio et al. 2011).  
Relevant model parameters include the minimum, break and maximum energy of the electron energy distribution,  $\gamma_{min}$,   
$\gamma_{b}$,   $\gamma_{max}$, respectively; the indices of the distribution below and above the break energy, $n_1$ and $n_2$, the magnetic field $B$; the particle density in the emitting region $K$, the size of the emitting region or blob $R$, and the Doppler boosting factor $\delta$ (Table~\ref{table:sedmodelparams}).  The model parameters are in general similar to those determined by  Lichti et al. (2008), Donnarumma et al. (2009), and Abdo et al. (2011) in their multiwavelength fits with a leptonic model.   Accommodating the softest LAT spectrum and the simultaneous TeV flux proved to be difficult (see Fig.~\ref{FigSED}).  While the discrepancy between the model and the TeV flux may be justified by the very high amplitude variability at these energies on short timescales, the softness of the LAT spectrum is unusual and may command substantial changes in our interpretation of this source.   Many authors have previously advocated more complex scenarios than the one we have adopted here, to account for the  MeV-GeV spectrum of Mkn~421, such as time-dependent and multicomponent models, lepto-hadronic jets, multizone models, electron energy laws with evolving spectra rather than steady-state spectra, inhomogeneous jets, spine-and-layer jet structure  (Krawczynski et al. 2001;  B{\l}a{\.z}ejowski et al. 2005; Ghisellini et al.  2005;  Aharonian et al. 2009; Potter \& Cotter 2013;  Mastichiadis et al. 2013; Asano et al. 2014).  We believe indeed that a time-dependent approach should primarily lead to a better description of the high-energy spectrum.
In view of  this, the large variation of the magnetic field  inferred from our modelling of the very soft LAT  spectrum of April 16 with respect to the  two following states may be reduced, when a  more  complex, but more plausible time-dependent description is implemented.

During the rising part of the first flare of this campaign the softer energy flux seems to lag behind  the higher energy flux with a delay that increases linearly with energy, but is formally not significant  (Fig.~\ref{FigDCF}).   Because of
this limitation and because the correlation cannot be tested on other parts of the light curves, we cannot use this to constrain  time-dependent models.
The complex multiwavelength variability observed in Mkn~421 during the present   outburst and previous ones  occurred and studied  over the past years indicates that the source behaviour may be dominated by different regimes of injection and cooling in different multiwavelength states, and thus must be monitored intensively during the transition phases to high states, in order to reconstruct  the physical conditions that produce  all regimes of variability.


\begin{table}
\caption{Model parameters of the multiwavelength energy distributions} 
\label{table:sedmodelparams}
\centering                       
\begin{tabular}{cccc}      
\hline\hline              
Parameter & 2013-Apr-16 (red$^{a}$) & 2013-Apr-17 (orange) & 2013-Apr-19 (black)  \\
\hline                 
$\gamma_{min}$         &    1000                          & 4000                          & 4000                             \\   
$\gamma_{b}$             &     31000                       &  $4.1 \times 10^5$     & $1.3 \times 10^5$        \\   
$\gamma_{max}$        &   $5 \times 10^5$          &   $2 \times 10^6$       & $1 \times 10^6$           \\   
$n_1$                          &      2                               &    2.5                          &    2.5                             \\    
$n_2$                          &    3.7                              &     4.9                         &    4.9                             \\   
$B$ (Gauss)                &    0.6                              &   0.038                        &    0.04                          \\   
$K$ (cm$^{-3}$)          &   1800                            &  $1.9 \times 10^5$      &  $1.1 \times 10^5$       \\   
$R$ (cm)                     &   $1.7 \times 10^{16}$   &  $1.3 \times 10^{16}$   &  $1.5 \times 10^{16}$  \\   
$\delta$                       &     10                              &   40                               &  40                              \\   
\hline  
\noalign{\smallskip}
\multicolumn{4}{l}{$^a$ Colours refer to the coding in  Figure~\ref{FigSED}. } \\
\end{tabular}                               
\end{table}


\begin{acknowledgements}

We thank Celia Sanchez, Marion Cadolle-Bel, Erik Kuulkers and Chris Winkler of the {\it INTEGRAL} Science Operation Centre for their 
assistance with the scheduling of the observations, Lucia Pavan for assistance with JEM-X data calibration, Imma Donnarumma for 
helpful discussion,  Eran Ofek for help with computational issues
and use of his astronomy \& astrophysics package for Matlab, and the anonymous referee for suggestions and comments that helped to improve the paper.  This work was partially supported by ASI/INAF contracts I/009/10/0 and I/088/06/0.  RH acknowledges GA CR grant 
102/09/0997.  This research has made use of the NASA/IPAC Extragalactic Database (NED) which is operated by the Jet Propulsion 
Laboratory, California Institute of Technology, under contract with the National Aeronautics and Space Administration. This paper is 
dedicated to the memory of our friend and colleague Paul Barr.

\end{acknowledgements}

\end{document}